\documentstyle [prb,aps,epsf]{revtex}
\begin{document}
\twocolumn[{
\widetext
\draft

\title{
Magnetoresistance and magnetic breakdown in the quasi-two-dimensional
conductors (BEDT-TTF)$_2$MHg(SCN)$_4$[M=K,Rb,Tl]
}

\author{Ross H. McKenzie,$^1$\cite{email}
G. J. Athas,$^2$ 
 J. S. Brooks,$^3$ R. G. Clark,$^1$ A. S. Dzurak,$^1$ R. Newbury,$^1$
 R. P. Starrett,$^1$\\
 A. Skougarevsky,$^{1}$ 
M. Tokumoto,$^4$ N.  Kinoshita,$^4$ T. Kinoshita,$^4$ and Y. Tanaka$^4$
}

\address{$^1$School of Physics and 
National Pulsed Magnet Laboratory, University of New
South Wales, Sydney 2052, Australia}
\address{
$^2$ Physics Department, Boston University,
Boston, MA 02215} 
\address{
$^3$ National High Magnetic Field Laboratory,
 Florida State University, Tallahassee, FL 32306}
\address{
$^4$ Electrotechnical Laboratory, Tsukuba, Ibaraki 305, Japan}

\date{Received 21 June 1996}
\maketitle
\mediumtext
\begin{abstract}
The magnetic field dependence of the resistance
of (BEDT-TTF)$_2$MHg(SCN)$_4$[M=K,Rb,Tl] in the
 density-wave phase  is explained
in terms of a simple model involving magnetic breakdown
and a reconstructed Fermi surface.
The theory is compared to measurements
in pulsed magnetic fields up to 51 T.
The value implied for the scattering time 
is consistent with independent determinations.
The energy gap associated with the density-wave phase
is deduced from the magnetic breakdown field.
Our results have important implications for the
phase diagram.
\\
\ \\
To appear in Physical Review B, Rapid Communications, September 15,
1996.
\end{abstract}


}]
\narrowtext

Conducting organic molecular crystals based on the BEDT-TTF and TMTSF 
molecules are novel low-dimensional electronic systems.\cite{ish,bro}
 The family (BEDT-TTF)$_2$MHg(SCN)$_4$[M=K,Rb,Tl,NH$_4$]
are of particular interest because they have 
a rich phase diagram
and coexisting
 quasi-one-dimensional and quasi-two-dimensional Fermi surfaces.
 Metallic, superconducting, and density-wave phases are possible,
 depending on temperature, pressure, magnetic field,
and anion type.\cite{bro}
At ambient pressure, the family with M= K,Rb,Tl undergo a transition
 from a metal
to a density-wave (DW) phase at a temperature $T_{DW} =$ 
8, 9, and 12 K, respectively.  There is
currently controversy as to whether this is a spin-density wave or a
charge-density wave.\cite{pratt,toyota,ath,mck2}
  This phase is destroyed above a magnetic field,
$H_k$, known as the kink field 
(for M=K,Tl, and Rb,  $H_k= $ 23, 27,  and 32 T,
 respectively).\cite{bro,brooks}

The purpose of this Rapid Communication is to present
new measurements of the field dependence
of the magnetoresistance up to 51 T and explain this             
dependence in terms of {\it magnetic breakdown} and  
a reconstructed Fermi surface in the DW phase.
The field dependence has the following features
(compare Figure 1). (1)  At low fields the resistance increases
rapidly up to $H_{max} \sim$  15T.  The maximum resistance is 
roughly an order
of magnitude larger than the zero field resistance.
(2)  The resistance then decreases with increasing field.
(3) Above about 30 T the background (non-oscillating)
 resistance saturates to a
value much larger than the zero-field resistance.
(4) At low temperatures
hysteresis is seen near the kink field.  This is
because destruction of the DW phase is a first order
transition at low temperatures.
(5) The maximum resistance increases and $H_{max}$
decreases as the temperature is lowered.
Measurements on poorer quality samples give smaller 
maximum resistance.\cite{osada2,sasaki3}
(6) As the angle between the field and the conducting planes is increased
$H_{max}$ increases\cite{ath,uji0,ath2,uji2}
 but $H_{k}$ does not vary.\cite{ath,ath2}

The measurements shown in Figure 1 were made at the Australian National
Pulsed Magnet Laboratory. \cite{cla}
Samples were studied in a top loading $^3$He refrigerator
and aligned so the magnetic field
was in the least-conducting direction (the ${\bf b}$ axis).
The voltage and current were also along the  ${\bf b}$ axis.
The magnet system was pulsed up to 51 T
 with a duration of 20 ms. Measurements were made with dc
constant current (80-200 $\mu$ A) sources and low noise,
differential pre-amplifiers. Pick-up from the dB/dt term was
never more than 50\% of the signal above 25 T.
The pick-up term was
eliminated from the data by  averaging forward and
reverse current traces.
A RuO$_2$
thermometer mounted within 5 mm of the sample was used to
monitor the temperature before and after each pulse. No
systematic changes in temperature were observed as a result of the pulse.
Preliminary data for a single temperature was
briefly reported elsewhere.\cite{mck5,bro3}
  Similar results have been obtained by
other groups on  the K and Tl salts
in fields up to  30T \cite{osada2,sasaki3,uji0,chen,kar2}
 and on K up to 50T.\cite{cau}

The room-temperature Fermi surface of 
(BEDT-TTF)$_2$MHg(SCN)$_4$[M=K,Rb,Tl]
 in the conducting plane, calculated within a tight binding
 model\cite{mori,rous}
 is shown in the inset of Figure 2.
  There is
a cylindrical or quasi-two-dimensional hole Fermi surface and a
quasi-one-dimensional electron Fermi surface consisting of two
 warped sheets. It
is believed that the nesting of the quasi-one-dimensional Fermi surface
is responsible for the formation of the DW phase.
The
DW introduces a new periodic potential with wavevector ${\bf Q}$ into
the
system  resulting in reconstruction of the quasi-two-dimensional
Fermi surface.
Two different reconstructions of the Fermi surface have
been proposed\cite{kar,uji1} and are described below.
We shall focus on the one shown in Figure 2,
{\it purely for reasons of calculational simplicity.}
We show here that if magnetic breakdown,
which causes the holes to return to
their original unreconstructed closed orbits, is
taken into account 
the complete field dependence of the resistance can be explained.
Similar results are expected for the second proposed Fermi
surface.\cite{prev}


In the DW phase the
large magnetoresistance oscillates
as the orientation of the
magnetic field relative
 to the most conducting planes is varied (angle-dependent
 magnetoresistance oscillations (AMRO)).\cite{bro}
To explain this effect a reconstructed Fermi surface consisting of
 of two open sheets and
many small ``lens'' orbits (Fig. 2)  
has been proposed.\cite{kar} 
The sheets give rise to a large magnetoresistance, except when
the current direction is perpendicular to the sheets.  At low fields
the magnetoresistance will increase quadratically with field.
This model has been used to give
a quantitative description of the AMRO 
for fields up to about 15 T.\cite{iye}
However, these calculations do not include
magnetic breakdown and cannot explain the decrease in resistance
with increasing fields above 15 T.

There are several problems with the Fermi
surface reconstruction shown in Figure 2.
The existence of open sheets                depends on a delicate
balance between the size and shape of the Fermi surface and the
direction of the DW wavevector.
There is experimental \cite{uji1,kovalev,haw}
and      theoretical\cite{gus} evidence
that the desired conditions are not met.
Uji {\it et al.}\cite{uji1} proposed an alternative reconstructed
Fermi surface with no open sheets. 
Compensated electron and hole pockets produce 
a large magnetoresistance which will be reduced
 by magnetic breakdown.\cite{ath}
Due to the above problems, Yoshioka\cite{yosh} proposed an 
explanation for the AMRO that does not require reconstruction of the
Fermi surface.\cite{str}

The effect of magnetic breakdown on
magnetoresistance has been considered
 in detail by Pippard\cite{pip} and Falicov and 
Sievert.\cite{fal}
  They quantitatively described the shape of the
magnetoresistance curves for zinc and magnesium,\cite{pip,sho}
 which are similar to those in Figure 1.
We have calculated the magnetoresistance for the model Fermi surface
shown in Figure 2 using the formalism of Falicov and Sievert.\cite{fal,sie}
The ratio of the resistance in a field $H$, $\rho(H)$, to
the zero field resistance, $\rho_0$, depends
on the dimensionless quantities $H/H_0$ and $eH_0 \tau/m^*$
where $\tau$ is the scattering time
(assumed to be the same at all points on the Fermi surface), $e$ is
the electronic charge, $m^*$ the effective mass, and $H_0$ is the magnetic
breakdown field\cite{pip}
\begin{equation}
H_0={ \pi  E_g^2 \over 2 e \hbar v_F^2 \sin 2 \theta}
\label{h0}
\end{equation}
where $E_g$ is the energy gap and $v_F$ is the Fermi velocity
and $\cos \theta= Q/2k_F$.\cite{angle}
The probability of magnetic breakdown occurring (i.e., a hole
tunnelling between the two pieces of Fermi surface) is $\exp(-H_0/H)$.
At high fields ($H \gg H_0$)
complete breakdown occurs, the holes simply perform closed orbits
and the resistance is independent of field and for 
the model Fermi surface (with $\theta = \pi/4$) \cite{fal}
\begin{equation}
\rho_\infty=\rho_0 \left(1 + {4 e H_0 \tau \over \pi m^*} \right).
\label{rhoinf}
\end{equation}
The holes experience an effective scattering rate \cite{fal}
$\tau^{-1}+ 4 e H_0/\pi m^*$
where the second term represents additional scattering due to magnetic
breakdown.\cite{hall}

Figure 3 shows the field dependence of the resistance 
for values of 
$e H_0 \tau / m^*$ ranging from 10 to 100.
The current is parallel to the open Fermi surface
and the field is perpendicular to the plane.  No
magneto-oscillations are present because the model is semiclassical. Note
the following features, all similar to that observed in 
(BEDT-TTF)$_2$MHg(SCN)$_4$[M=K,Rb,Tl].
  (1) For low
fields the resistance increases quadratically with field.
(2)  There is a maximum at a field $H_{max}$.
(3)  Above about $0.8 H_0$ the resistance depends weakly
on the field and on the scattering rate.
(4)  As the scattering rate decreases the maximum value of the resistance
increases and $H_{max}$ decreases.

It should be noted that the current orientation
in our calculation is {\it not} the same as
in the experiment. In the experiment 
the current and field were set parallel to the
least conducting direction, as others have done,
because this produces a large signal to noise ratio.
In such a configuration no Lorentz force acts on the
electrons and so no classical magnetoresistance
and no oscillations are expected. Yet, for reasons 
that are not understood,\cite{ath} 
the data is similar to that seen when the
current is in the most conducting plane.\cite{osada2,sasaki3,uji0,sasaki}

Comparing     our data for Tl to the theory gives       
values for $\tau$  and $H_0$ of $(3 \pm 2) \times 10^{-12}$ sec and 
60 $\pm$ 20 T, respectively.
The value of $\tau$ corresponds to a Dingle temperature
of 0.4 $\pm$ 0.3 K. This value is comparable to values of about
0.2 K deduced from the field dependence of SdH and dHvA
oscillations {\it above} $H_K$ for the K salt.\cite{harr} This value is much
smaller than the values of about 3-4 K deduced
from the field dependence of the oscillations
{\it below} $H_K$.\cite{harr} This may be reasonable because
the field dependence of the closed hole  orbit
(also known as the $\alpha$ orbit)
below $H_K$ will be dominated by magnetic
breakdown and not scattering.\cite{est}

The temperature dependence of the magnetoresistance might
appear to be due to the temperature dependence of the scattering 
rate. If so    the scattering rate
in the Tl salt should change
 by a factor of about two  as the temperature changes
from 0.36 to 4.4 K. However, no such change is observed in
the zero field resistance. \cite{sasaki2}

The deduced value of $H_0$  and
 (\ref{h0})
 gives a value for $E_g$ of 10 $\pm$ 2 meV.\cite{mass}
It is important to note that the {\it same}
 periodic potential (due to the DW wave) 
reconstructs the hole Fermi surface
and produces an energy gap $E_1$
on the quasi-one-dimensional electron Fermi surface.
Elementary band theory\cite{ash} implies $E_1=E_g$.
As far as we are aware  $E_1$   has not 
been determined previously.
A rough estimate of this gap can be made
by noting that for a quasi-one-dimensional system (with no
coexisting two-dimensional Fermi surface) mean-field theory implies
$E_1 = 3.52 k_B T_{DW}$.  A transition
temperature of $T_{DW} = 9 $ K gives $E_1 = 3$
  meV.  However, in typical
quasi-one-dimensional materials the gap is actually two to five times
that predicted by this    relation (see Table II in Ref.
 \onlinecite{mck}),
probably due to fluctuations reducing the transition temperature.
Hence the value we deduce for the 
breakdown field is quite reasonable.
For the Rb salt we deduce a slightly larger value of $H_0,$
and thus $E_1 $,   consistent with
the trend in transition temperatures (9 K versus 12 K).

That we can describe the field dependence of the resistance 
using the magnetic breakdown model applied to the reconstructed
Fermi surface has important implications for the phase diagram and what
one deduces from magnetoresistance measurements.  Within our framework
the transition at the kink field represents only a small change in the
magnetoresistance.  In contrast, for the
Tl salt it has been suggested that because the
resistance decreases between $H_{max}$ and 
$H_k$ this field region represents
a different phase.\cite{sasaki3,sasaki}
  Also, it has been suggested that the absence of 
AMRO above $H_k$ denotes destruction of the reconstructed Fermi
 surface.\cite{cau}
However, within our model this may not be the case: the Fermi surface
may still be reconstructed but due to magnetic breakdown the open
Fermi surface has little effect on the resistance.
The question of the nature of the high field phase will be considered
in more detail elsewhere.\cite{mck2}

In conclusion,
 we have presented measurements of the field and
temperature dependence of the  resistance of 
(BEDT-TTF)$_2$MHg(SCN)$_4$[M=Rb,Tl]
 up to 51T and shown how the field dependence 
can be explained in terms of magnetic breakdown and a 
reconstructed Fermi surface  in the density-wave phase.
Our successful explanation has important implications for the phase
diagram.
It is not necessary to assume that there is
a new phase between $H_{max}$ and $H_k$, and
the high field phase may not be the same as the 
zero field metallic phase.



Work at UNSW was supported by the Australian Research Council.
GJA and JSB were supported in part by NSF grant DMR 92-14889.
We  thank P. M. Sievert, J. Singleton, S. Uji, 
and T. Ziman for helpful discussions.
We thank M. V. Kartovsnik for 
sending us a well-characterized sample of
(BEDT-TTF)$_2$TlHg(SCN)$_4$.
Figure 2 was produced by D. Scarratt.
We thank T. Ziman and E. Canadell for providing
the data for the inset of Figure 2.


\vskip -2.7cm
\begin{figure}
\centerline{\epsfxsize=13.3cm 
 \epsfbox{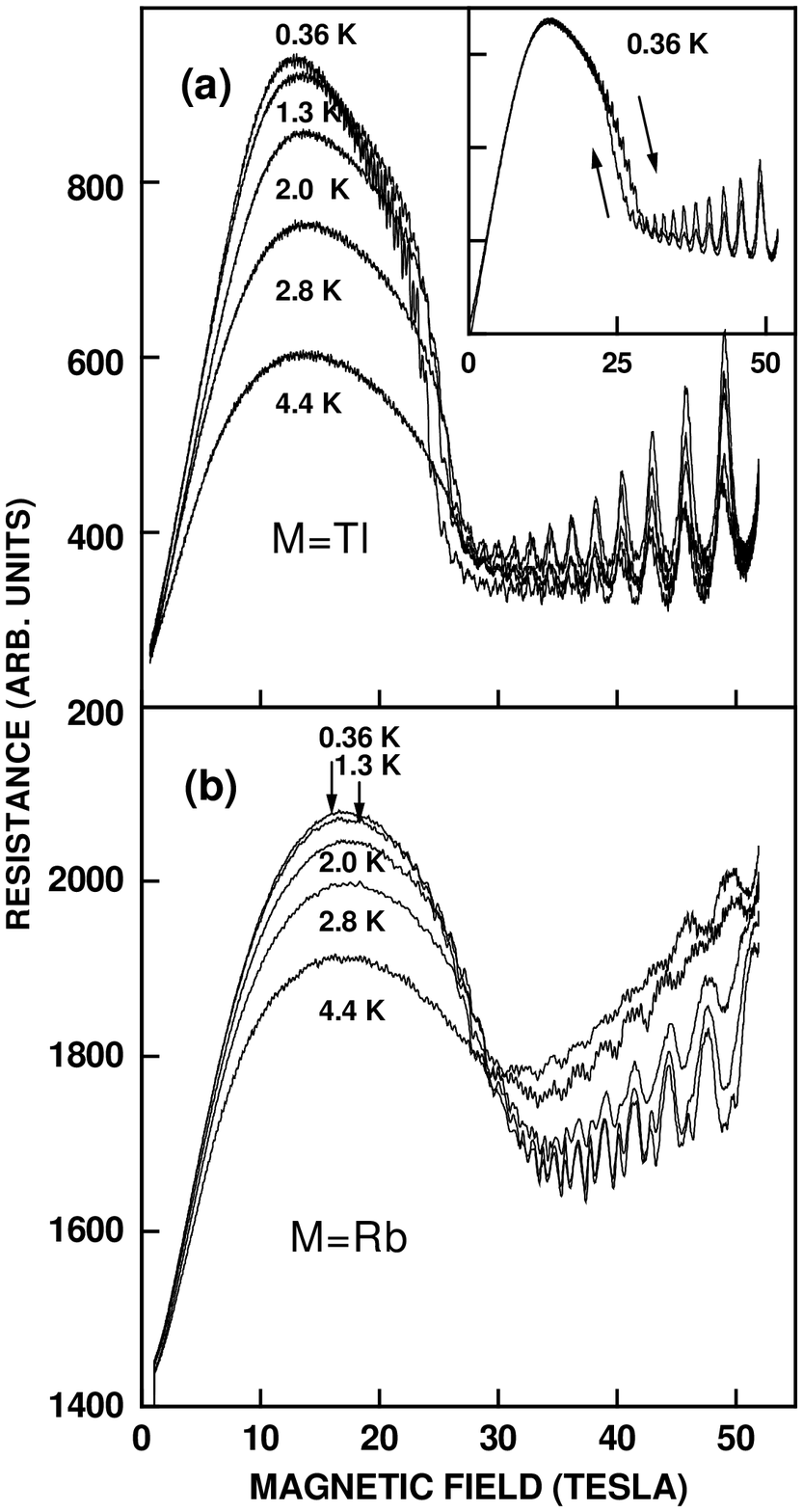}}
\vskip -3.7cm
\caption{
Magnetic field dependence of the resistance of 
(BEDT-TTF)$_2$MHg(SCN)$_4$ 
 at different temperatures for (a) M=Tl and (b) M=Rb.  The pulsed
magnetic field and the current direction were parallel to the
least-conducting direction.  Note that the resistance increases
rapidly up to about 15T, then decreases until about 30T.
The inset of (a) shows  two curves 
corresponding to up and down sweeps of the magnetic field.
They do not coincide near 27T (the "kink field") due to hysteresis
associated with the first order transition there.
For clarity only down sweeps are shown in the main
Figure.
The measurements on Tl were four terminal and
those on Rb were two terminal with a large 
contact resistance.
\label{fig1}}
\end{figure}

\begin{figure}
\
\centerline{
\epsfxsize=10.0cm 
 \epsfbox{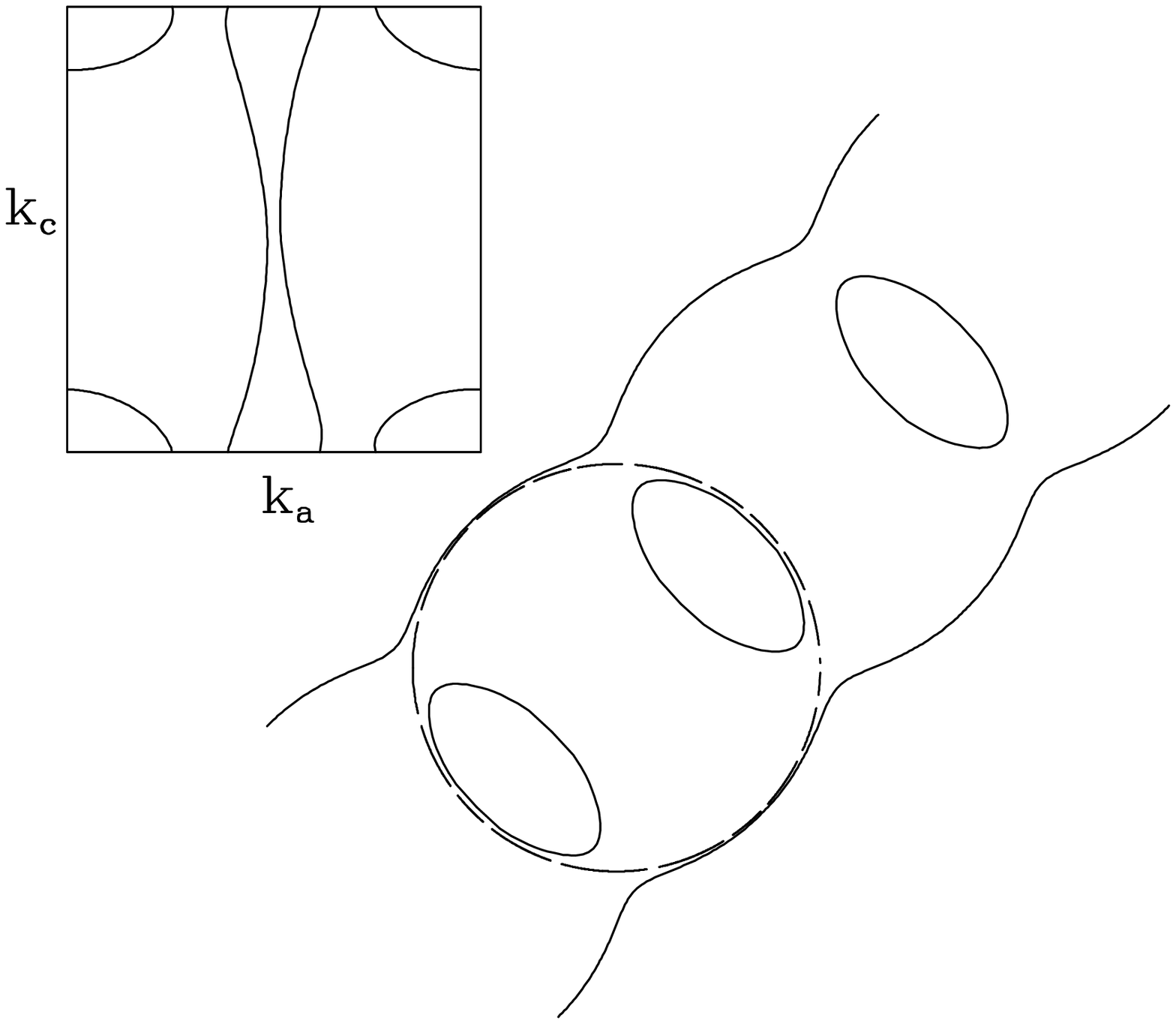}}
\caption{
One  possible reconstruction of
 the Fermi surface by the periodic potential due to 
the density wave.
The inset shows  the calculated Fermi surface for the Tl
salt \protect\cite{rous,gus}
 at room temperature.
It consists of quasi-two-dimensional cylinders for holes
 and quasi-one-dimensional open sheets for electrons. 
The main figure shows the
 reconstructed hole Fermi surface used in our calculations. It
now comprises open orbits and closed orbits.  The former produce a
large magnetoresistance at low fields.  At high fields magnetic
breakdown results in only closed orbits (dashed lines).
  The open electron Fermi surface shown in the inset
disappears due to the opening of an energy gap.
\label{fig2}}
\end{figure}

\vskip -0.7cm
\begin{figure}
\centerline{\epsfxsize=7.0cm  \epsfbox{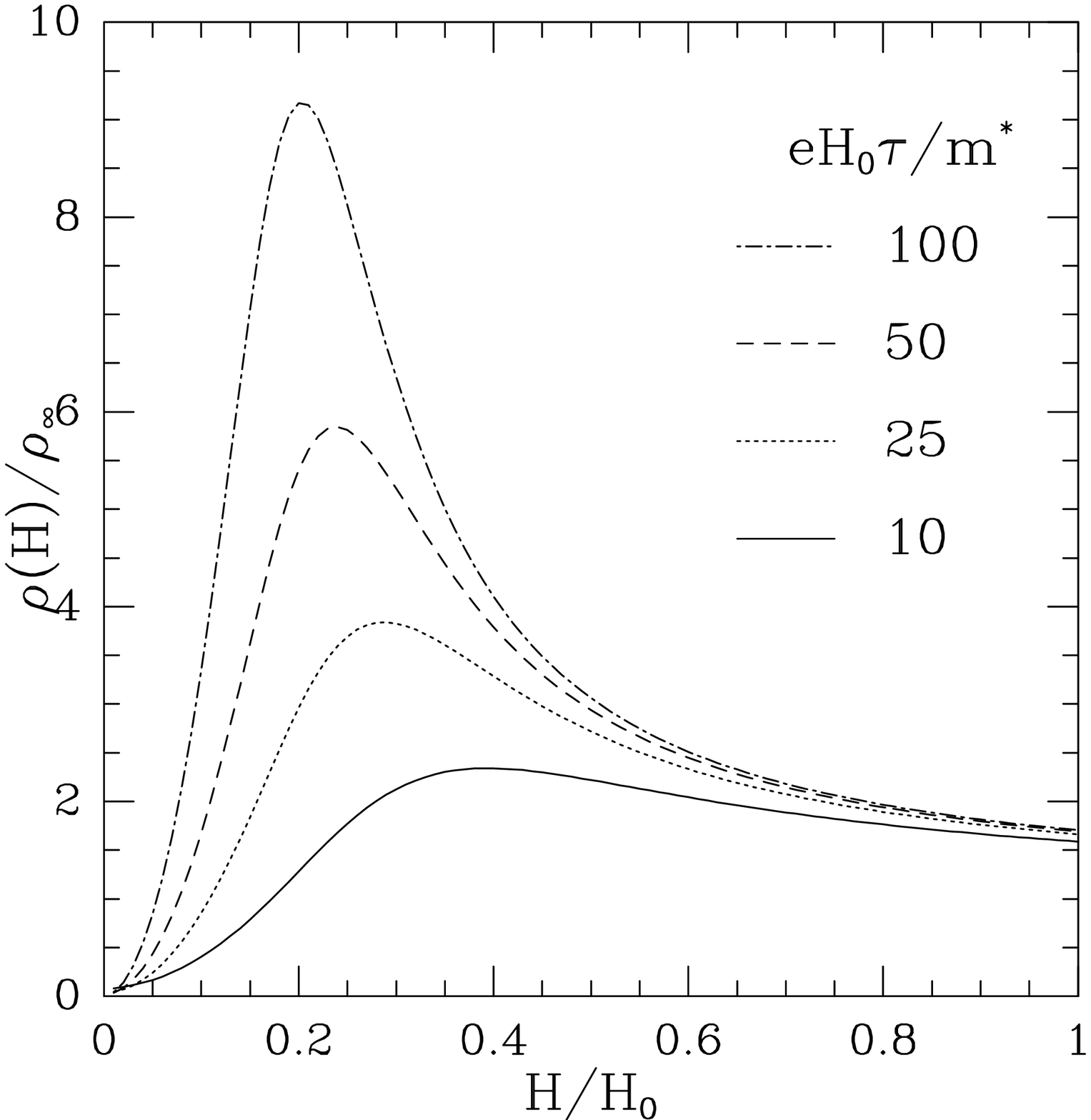}}
\vskip 0.3cm
 \caption{
Magnetic field dependence of the resistance for the Falicov-Sievert
model with the Fermi surface shown in Figure 2.  The calculation is
for a 
field perpendicular to the plane and the current parallel to the open
sheet of the reconstructed  Fermi surface.
The upper curves correspond to larger scattering times
$\tau$, i.e., lower temperatures
or higher quality samples.  The
magnetic field is normalised to the magnetic breakdown field 
$H_0$ defined in Eq. (\protect\ref{h0}).
The resistance is normalised to its value at high fields
given by Eq. (\protect\ref{rhoinf}).
A similar field dependence is expected for the 
alternative Fermi surface proposed by Uji {\it et al.}
\protect\cite{uji1}
\label{fig3}}
\end{figure}


\begin{references}
\bibitem{prev}
It has previously been suggested \cite{osada2,uji0}
that the field dependence can be explained in
terms of magnetic breakdown. However, this
earlier work assumed that the breakdown was
between the open and closed Fermi surfaces in the
inset of Figure 2
(producing what is known as the $\beta$ orbit)
and involves a breakdown field two orders of
magnitude larger than observed.
\bibitem{angle}
For simplicity we assume $\theta= \pi /4.$
Figure 2 in Ref. \protect\onlinecite{fal}
 shows that varying $\theta$ produces only small changes
in the shape of the magnetoresistance curve.
\bibitem{hall}
For the model Fermi surface considered here
magnetic breakdown has no effect on the Hall resistivity.  For all fields
it is given by $H/ne$ where $n$ is the density of holes. 
Note that if the sample purity is sufficiently great 
that $ eH_0 \tau \gg m^*$  then
$ \rho_\infty \sim H_0/ne $ and
the ratio of the resistance to the Hall
resistance will be roughly $ H_0/H.$
Hall measurements\protect\cite{sasaki2}
 are consistent with this if $ H_0 \sim 20 $ T.
\bibitem{est}
If so then the amplitude of the oscillations
$\sim \exp (-2 H_0/H)$ (Ref. \protect\onlinecite{sho}, p.343 ff.). 
This means the Dingle temperature
of 3-4 K (Ref. \protect\onlinecite{harr})
 corresponds to $H_0 \simeq $ 60-80T .
Although the consistency of this
with our independent estimate of $H_0$ is appealing,
this argument would also predict the same field
dependence for the first and second harmonic,
contrary to experiment.\protect\cite{harr}
\bibitem{mass}
The Fermi velocity was  estimated from
the Fermi wave vector $k_F$
calculated from the  area of the hole Fermi surface
implied by the frequency (670 T)\cite{mck5}
 of Shubnikov de Haas oscillations.\cite{sho}
We used an effective mass $m^* = 2.7 m_e$, given by the temperature
dependence of the SdH oscillations.\protect\cite{harr}
\end{references}
\end{document}